\documentclass[aps,prl,twocolumn,superscriptaddress,notitlepage,nofootinbib]{revtex4-1}
\usepackage{amsmath,amssymb,amsfonts,bm}
\usepackage{graphicx,color}
\usepackage[colorlinks=true,linkcolor=blue,plainpages=false]{hyperref}
 \usepackage{relsize}
 
\newcommand{\md}{\mathrm{d}}

\begin{document}
\title{Medium-Assisted Enhancement of Exotic Hadron Production 
\\ 
from Small to Large Colliding Systems}

\author{Yu Guo}
\affiliation{Guangdong Provincial Key Laboratory of Nuclear
Science, Institute of Quantum Matter, South China Normal University, Guangzhou 510006, China}
\affiliation{Guangdong-Hong Kong Joint Laboratory of Quantum
Matter, Southern Nuclear Science Computing Center, South China Normal University, Guangzhou 510006, China}

\author{Xingyu Guo}
\email{guoxy@m.scnu.edu.cn}
\affiliation{Guangdong Provincial Key Laboratory of Nuclear
Science, Institute of Quantum Matter, South China Normal University, Guangzhou 510006, China}
\affiliation{Guangdong-Hong Kong Joint Laboratory of Quantum
Matter, Southern Nuclear Science Computing Center, South China Normal University, Guangzhou 510006, China}

\author{Jinfeng Liao}
\email{liaoji@indiana.edu}
\affiliation{Physics Department and Center for Exploration of Energy and Matter,
Indiana University, 2401 N Milo B. Sampson Lane, Bloomington, IN 47408, USA}

\author{Enke Wang}
\email{wangek@scnu.edu.cn}
\affiliation{Guangdong Provincial Key Laboratory of Nuclear
Science, Institute of Quantum Matter, South China Normal University, Guangzhou 510006, China}
\affiliation{Guangdong-Hong Kong Joint Laboratory of Quantum
Matter, Southern Nuclear Science Computing Center, South China Normal University, Guangzhou 510006, China}

\author{Hongxi Xing}
\email{hxing@m.scnu.edu.cn}
\affiliation{Guangdong Provincial Key Laboratory of Nuclear
Science, Institute of Quantum Matter, South China Normal University, Guangzhou 510006, China}
\affiliation{Guangdong-Hong Kong Joint Laboratory of Quantum
Matter, Southern Nuclear Science Computing Center, South China Normal University, Guangzhou 510006, China}

\begin{abstract} 
Studies of exotic hadrons such as the $\chi_{c1} (3872)$ state provide crucial insights into the fundamental force governing the strong interaction dynamics, with an emerging new frontier to investigate their production in high energy collisions where a partonic medium is present. Latest experimental measurements from the Large Hadron Collider show an intriguing evolution pattern of the $\chi_{c1} (3872)$-to-$\psi(2S)$ yield ratio from proton-proton collisions with increasing multiplicities toward proton-lead and lead-lead collisions. Here we propose a novel mechanism of medium-assisted enhancement for the $\chi_{c1} (3872)$ production, which competes with the more conventional absorption-induced suppression and results in a non-monotonic trend from small to large colliding systems. Realistic simulations from this model offer the first quantitative description of all available data. Predictions are made for the centrality dependence of this observable in PbPb collisions as well as for its system size dependence from OO and ArAr to XeXe and PbPb collisions. In both cases, a non-monotonic behavior emerges as the imprint of the competition between enhancement and suppression and can be readily tested by future data.  
\end{abstract}
\maketitle

{\it Introduction}.
The overwhelming majority of the energy and mass in the visible component of our universe comes from the strongly interacting elementary particles, or hadrons, such as protons, neutrons and pions. According to the fundamental theory of elementary particles known as the Standard Model, these hadrons are themselves made from quarks and antiquarks whose interactions are governed by a basic theory of strong interaction --- the Quantum Chromodynamics (QCD). While the QCD equations are known, their full consequences are difficult to decipher. One of the outstanding challenges is about the so-called exotic hadrons, whose quark/antiquark configurations do not follow the established normal patterns of three quarks forming a baryon (such as the protons and neutrons) and a quark-anti-quark pair forming a meson (like pions). 

A most notable example of such states is the $\chi_{c1} (3872)$ particle, also commonly known as $X(3872)$, first discovered by the Belle experiment\cite{Belle:2003nnu} in 2003. Its quark content consists of a pair of charm and anti-charm quarks as well as another pair of light flavor quark and anti-quark, i.e. $c\bar{c}q\bar{q}$. Subsequently extensive efforts~\cite{CDF:2003cab,D0:2004zmu,LHCb:2020xds,LHCb:2020fvo,BESIII:2020nbj,BESIII:2013fnz,CMS:2021znk,CMS:2021znk,BaBar:2019hzd,Belle:2017psv,Li:2019kpj,BESIII:2022kow,Belle:2022puc,LHCb:2022bly,BaBar:2008flx,Belle:2011wdj,LHCb:2014jvf} have been made to measure its quantum numbers as well as to find many other candidates of exotic hadrons. There are many open questions in this very active research frontier: What kinds of exotic hadrons could exist? What are the internal structures and properties of them? What are their production and decay mechanism?  

While properties of  $\chi_{c1} (3872)$ have been studied at multiple colliders, physicists remain puzzled by the nature of this particle despite 20 years past its initial discovery, with multiple interpretations being proposed for its internal structures, such as a   large-size hadronic molecule versus  a compact-size tetraquark state.
See recent reviews in e.g. \cite{Brambilla:2010cs,Esposito:2014rxa,Briceno:2015rlt,Shepherd:2016dni,Hosaka:2016pey,Esposito:2016noz,Lebed:2016hpi,Olsen:2017bmm,Guo:2017jvc,Wu:2022ftm,Chen:2022asf}.



\begin{figure}[!hbt]
	\begin{center}
		\includegraphics[width=3.3in]{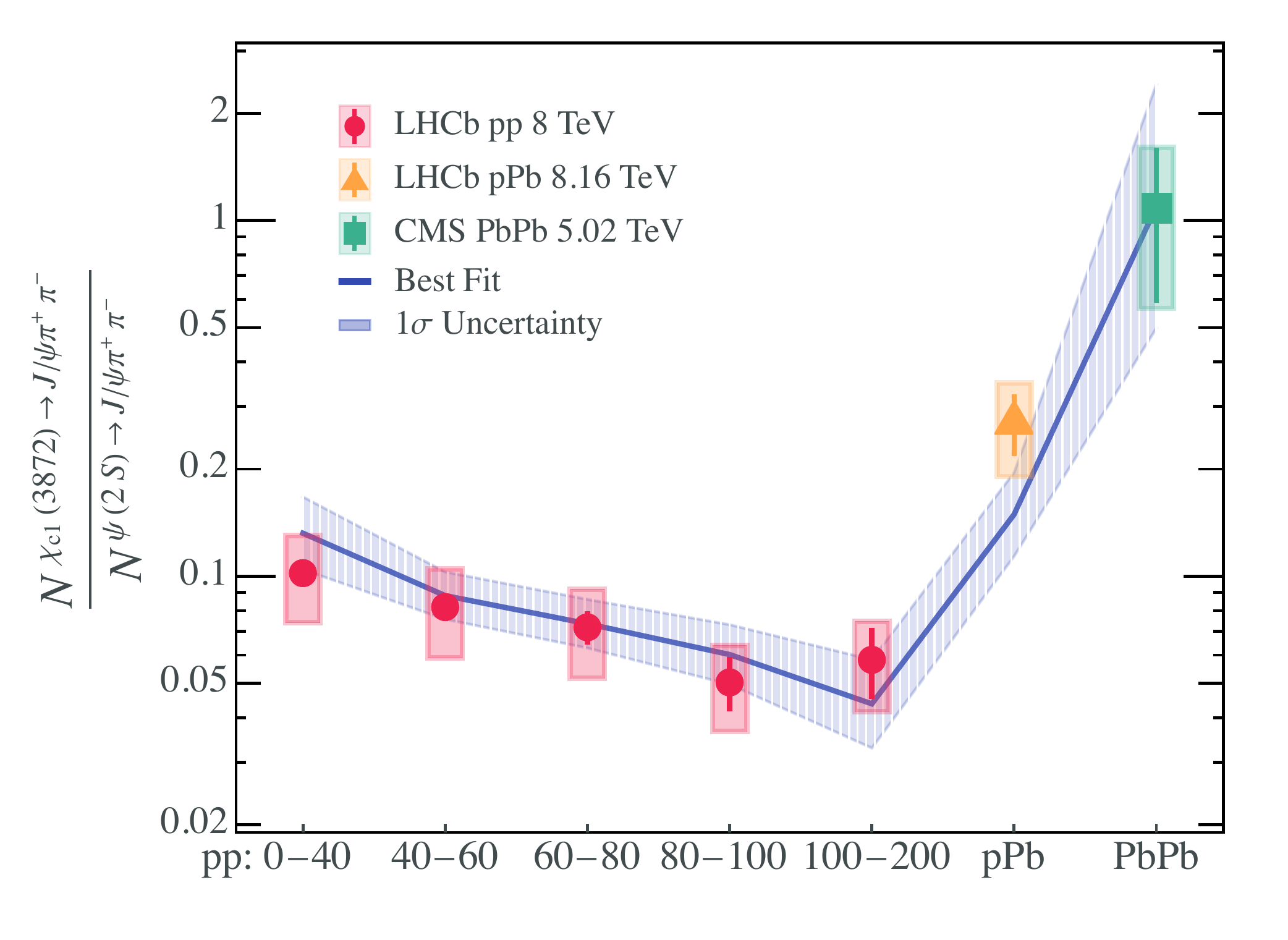}
		\caption{ A comparison of the $\chi_{c1} (3872)$ yield relative to $\psi(2S)$ between model simulation results (blue curve) and experimental data from LHCb pp collisions at $\sqrt{S_{NN}}=8$ TeV (red circle),   LHCb preliminary pPb collisions at $\sqrt{S_{NN}}=8.16$ TeV (orange triangle) and  CMS PbPb collisions at $\sqrt{S_{NN}}=5.02$ TeV (green box)~\cite{CMS:2021znk,LHCb:2020sey,LHCb:2022ixc}. 		
The model parameters are determined from the global fitting analysis with the blue band showing the $1\sigma$ level uncertainty. (See text for details.)}
		\vspace{-0.5cm}
		\label{fig_2}
	\end{center}
\end{figure}

A new avenue of investigating exotic hadrons has recently emerged and rapidly developing, namely to study their formation in high energy hadron and nuclear collisions where a partonic medium is present. In such collisions, a fireball with many light flavor quarks/antiquarks (on the order of hundreds to thousands depending on colliding systems) is created together with a considerable number of charm quarks/antiquarks. This provides an ideal environment for creating heavy flavor exotic states and probing their properties, as demonstrated in the latest theoretical works~\cite{ExHIC:2010gcb,ExHIC:2011say,ExHIC:2017smd,Hong:2018mpk,Fontoura:2019opw,Zhang:2020dwn,Esposito:2020ywk,Braaten:2020iqw,Wu:2020zbx,Abreu:2020jsl,Hu:2021gdg,Chen:2021akx,Albaladejo:2021cxj,Xu:2021drf,Abreu:2021jwm,Jin:2021cxj,Abreu:2022lfy,Braaten:2022qag,Yun:2022evm,Montana:2022inz,Hu:2023hrn}.   
 Most importantly, experimental measurements of $\chi_{c1} (3872)$ production in these collisions have started to arrive in the last few years, including LHCb data from high multiplicity proton-proton (pp) collisions~\cite{LHCb:2020sey} and proton-lead (pPb) collisions~\cite{LHCb:2022ixc} as well as CMS data from lead-lead (PbPb) collisions~\cite{CMS:2021znk} at the Large Hadron Collider (LHC). Already this first batch of empirical information shows an unusual pattern of the partonic medium's influence on the $\chi_{c1} (3872)$ yield with respect to the yield of another particle called  $\psi(2S)$ which is a normal hadronic state serving as a benchmark for comparison, by virtue of its similar heavy flavor content and decay channel ($J/\Psi\pi\pi$)  as well as close mass value to the exotic $\chi_{c1} (3872)$. (These data points are shown in Fig.~\ref{fig_2}.) The LHCb pp results suggest the yield ratio of $\chi_{c1} (3872)$ relative to $\psi(2S)$ decreases with increasing event multiplicity,  which would hint at a suppression effect due to the medium. On the other hand,  the LHCb pPb results and the CMS PbPb results, for which the generated medium is expected to be larger in terms of parton density and system size as compared with pp collisions, show a strong increase in this observable, which would indicate an opposite trend to the pp results. So far there has lacked a consistent explanation that reconciles this intriguing behavior of $\chi_{c1} (3872)$ production from small to large colliding systems. 
 


In this Letter, we present a phenomenological model for the partonic medium attenuation effects on the production of  $\chi_{c1} (3872)$  in high energy hadron and nuclear collisions. 
In particular, a novel mechanism of medium-assisted enhancement effect will be proposed which competes with the more conventional absorption-induced suppression effect.
Based on this important feature, it will first be demonstrated qualitatively how the competition leads to a nontrival pattern in the yield ratio of  $\chi_{c1} (3872)$ relative to the $\psi(2S)$ while the partonic medium evolves from the smaller to the larger systems. We will then utilize realistic simulations to show how such a model offers the first quantitative description of all available experimental data. Predictions will also be made for observables that can be verified in the future.

{\it Method.}
In the high $p_T$ region where recent CMS and LHCb measurements were made, the production of $\psi(2S)$ and $\chi_{c1} (3872)$ should dominantly come from virtual  $c\bar{c}$ pairs generated in the initial hard scatterings. Suppose the number of such pairs that would eventually turn into $\psi(2S)$ and $\chi_{c1} (3872)$, in the absence of any medium effect, would be $N_{\psi(2S)}$ and $N_X$ respectively.  However, in nucleus-nucleus (AA) or high-multiplicity pp and proton-nucleus (pA) collisions, 
these pairs will need to first travel through the created partonic medium before producing those final hadrons. The influence of the medium on the evolution of such $c\bar{c}$ pairs is the focus of our analysis.  

The first important effect is the medium absorption. Random collisions with quarks and gluons from the medium result in the dissociation of the correlated co-moving $c\bar{c}$ pair, which is akin to the well-known $J/\Psi$ suppression as well as jet quenching that have been observed in AA collisions~\cite{Rapp:2008tf,Rapp:2023}. We model this effect as the geometric absorption along the in-medium path of a $c\bar{c}$ pair: 
\begin{eqnarray} 
\frac{\md N_i}{\md x}=-\alpha_i n(x) N_i,
\end{eqnarray}
where $i\to \psi(2S),\chi_{c1} (3872)$. The $n(x)$ is the local parton density of the medium along the path of a surviving $c\bar{c}$ pair. The coefficient $\alpha_i$ describes the likelihood of a given state to be dissociated, with the dimension of a cross-section. While the precise internal structure of $\chi_{c1}$ is unknown, given the higher mass of $\chi_{c1}$ and its near-threshold nature, its binding energy is plausibly smaller than that of $\psi(2S)$. Thus one expects a stronger absorption effect for $\chi_{c1}$, similarly to the sequential suppression studied for normal charmonium and bottomonium states~\cite{Karsch:2005nk,CMS:2012gvv,STAR:2022rpk,Rapp:2008tf,Rapp:2023,Du:2015wha}.
As is typically done in geometric models for jet energy loss or for charmonium suppression, one can evaluate the overall suppression by first integrating the above equation along any given path, then averaging over all possible in-medium paths, and finally averaging over collision events. This leads to the following expression for   the suppression factor of $\psi(2S)$: 
\begin{eqnarray}
R^{\psi(2S)}=\langle\langle e^{-\alpha_{\psi(2S)} \int_{\mathrm{path}}n(x)\md x}\rangle \rangle 
\end{eqnarray} 
where the notation $\langle\langle ... \rangle\rangle$ means $\langle\langle ... \rangle_{\mathrm{path}}\rangle_{\mathrm{event}} $. 
We note that such a suppression effect applies similarly to the $\chi_{c1} (3872)$ production. 


For $\chi_{c1} (3872)$, however, there is another medium effect that can actually help enhance its production. In addition to the $c\bar{c}$, the formation of $\chi_{c1} (3872)$ requires two light quarks/antiquarks. Scatterings with the partonic medium, which serves as a reservoir of numerous light quarks/antiquarks,  could lead to ``picking up'' of light quarks/antiquarks which then co-move with the $c\bar{c}$ pair. This enhances the probability to form the $\chi_{c1} (3872)$ state in the end. One could consider this as a two-step process, in which the $c\bar{c}$ pair picks up the first needed light parton and subsequently a second needed light parton. Therefore one can model such a \emph{medium-assisted enhancement} effect as follows:
\begin{eqnarray}
\frac{\md N_{\chi_{c1}  }}{\md x}=\beta_{X} n(x)\left[\int_0^x \beta_X n(y)\md y\right] N_X,
\end{eqnarray}
where $\beta_X$ is a parameter characterizing the probability of picking up a single light parton, which also has the dimension of a cross-section. An important feature of this effect is that it scales as square power of the medium parton density. The enhancement mechanism arises from utilizing light partons from medium for producing the exotic state along in-medium path of $c \bar{c}$ pairs, which differs from the regeneration mechanism proposed for the production of normal quarkonium states~\cite{Grandchamp:2002wp,Du:2015wha,Du:2017qkv,Zhou:2014kka,Rapp:2008tf}.

Combining this enhancement together with the previous suppression effect, one obtains:  
\begin{eqnarray}
R^X=\langle\langle e^{ \int_{\mathrm{path}}[-\alpha_X n(x)+\beta_X^2n(x) \int_0^x n(y)\md y] \md x}\rangle \rangle \, .
\end{eqnarray} 
The two parameters $\alpha_X$ and $\beta_X$ quantify the suppression and enhancement due to dynamical interactions with medium. Their values should be sensitive to whether the  $\chi_{c1} (3872)$ is a hadronic molecule or a tetraquark state. Thus extracting them from experimental data can help shed light on  the microscopic structure of $\chi_{c1} (3872)$.

Now we can compare the production of $\chi_{c1} (3872)$ relative to $\psi(2S)$. This is quantified by the ratio of their baseline pp production cross-section, modulated by their respective suppression/enhancement effects along the in-medium paths:
\begin{eqnarray}
\frac{N^X}{N^{\psi(2S)}} &=& \frac{\sigma^X_{pp}}{\sigma^{\psi(2S)}_{pp}}\times\frac{R^X}{R^{\psi(2S)}}\nonumber\\
&\approx &  \frac{\sigma^X_{pp}}{\sigma^{\psi(2S)}_{pp}} \times \mathcal{R}_{med.},  \\
\mathcal{R}_{med.}  &\equiv& \langle\langle e^{ \int_{\mathrm{path}}[-(\alpha_X-\alpha_{\psi(2S)}) n(x)+\beta_X^2 n(x)\int_0^x n(y)\md y] \md x}\rangle\rangle .\nonumber\\
\end{eqnarray}
In the above, the pp baseline $\sigma^X_{pp}/\sigma^{\psi(2S)}_{pp}$ could be inferred from experimental data. We will focus on analyzing the medium attenuation factor $\mathcal{R}_{med.}$.  Clearly, $\mathcal{R}_{med.} > 1$ implies an overall medium enhancement while 
$\mathcal{R}_{med.} < 1$ means an overall medium suppression for the $\chi_{c1} (3872)$ production relative to the $\psi(2S)$. While $\chi_{c1} (3872)$ and $\psi(2S)$ may have different formation time, we focus on relatively high $p_T$ region where both states presumably form at late time and assume the same in-medium paths for the initial $c \bar{c}$ pairs.

Let us first examine the qualitative feature of the partonic medium effect. Since $\alpha_X > \alpha_{\psi(2S)}$ and thus $\alpha_X - \alpha_{\psi(2S)}>0$, the first term in the exponential of $\mathcal{R}_{med.}$  is a suppression term. Its contribution grows linearly with the medium parton density and the path length. The second term is an enhancement term and its contribution grows quadratically with the parton density and path length. As a result, for relatively low medium density and/or small medium size, the first term will dominate and therefore the overall medium effect would be a suppression of $\chi_{c1} (3872)$ relative to $\psi(2S)$. On the other hand, for relatively high medium density and large medium size, the second term will dominate and therefore the overall medium effect would be an enhancement, instead. This nonlinear feature, arising from the competition between suppression and enhancement, points to a non-monotonic evolution of medium effect that can help explain and provide a unified interpretation of the recent measurements by both LHCb and CMS from small to large colliding systems. 
For a simple illustration, let us assume that the average medium effect can be approximated by an average parton density $\bar{n}$ and average path length $\bar{L}$. Further defining a  medium thickness parameter $\bar{W}\equiv \bar{n}\cdot \bar{L}$, the factor $\mathcal{R}_{med.}$ can be simplified as
\begin{eqnarray}
\mathcal{R}_{med.} =   e^{ \left [- \left ( \alpha_X-\alpha_{\psi(2S)} \right ) \bar{W}  +\frac{1}{2}\beta^2_X \bar{W}^2 \right ]}.
\end{eqnarray} 
The above result clearly suggests that changing from pp with increasing multiplicities through pA eventually to AA collisions, the medium thickness $\bar{W}$ will monotonically increase so that the medium attenuation factor $\mathcal{R}_{med.} $ should first decrease and then increase, for which a minimum would occur at $\bar{W}= \bar{n}\bar{L}=\frac{\alpha_X-\alpha_{\psi(2S)}}{\beta^2_X}$. 

\begin{figure}[!hbt]
	\begin{center}
		\includegraphics[width=3.3in]{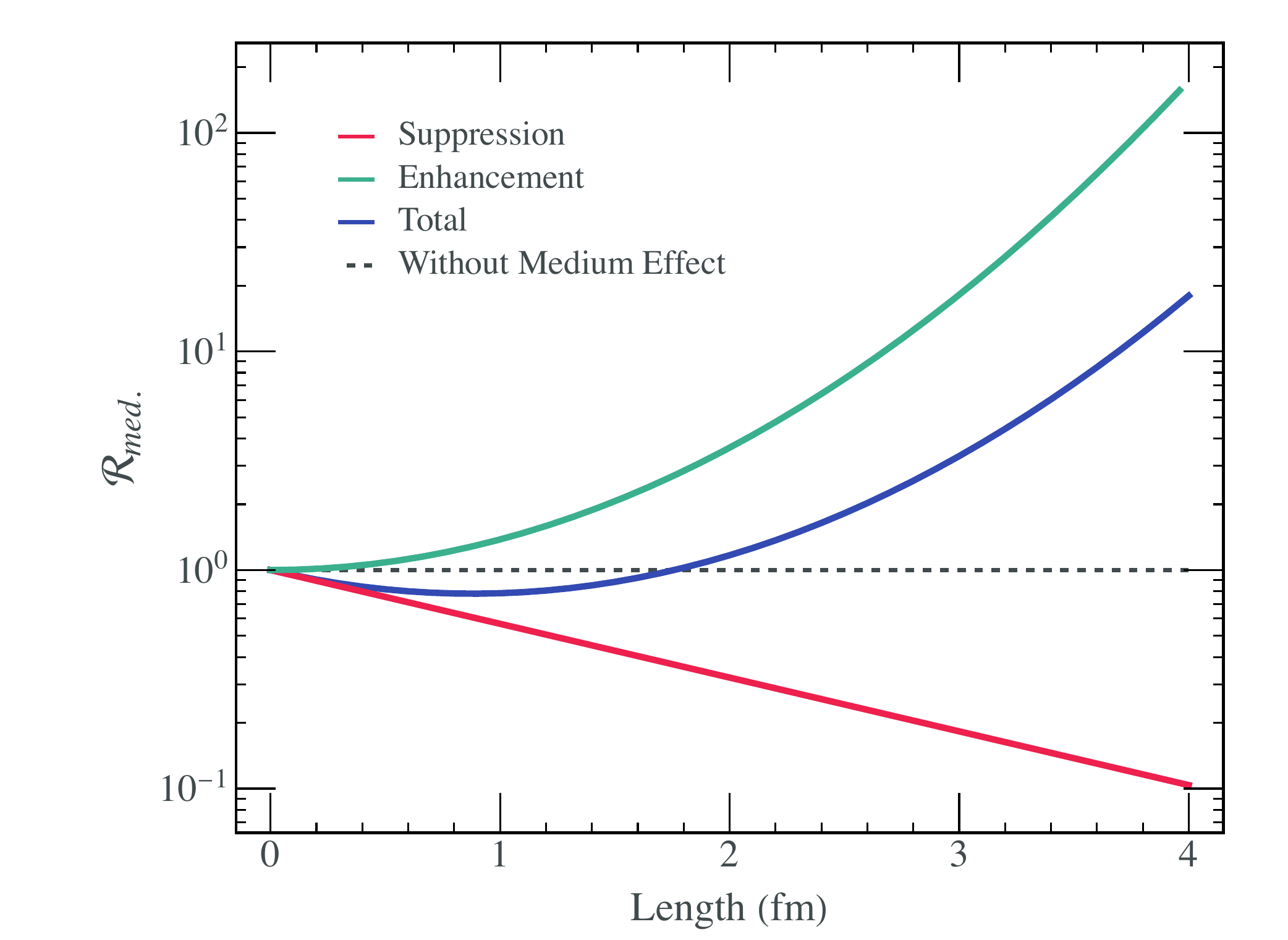}
		\caption{
Individual contributions from absorption-induced suppression (blue) and medium-assisted enhancement (red) as well as the overall $\mathcal{R}_{med.}$ (black) are plotted as functions of the QGP ``brick'' length (in fm unit). The QGP is set at a temperature of $360\rm MeV$ corresponding to an entropy density of $105\rm fm^{-3}$. 		
The dashed line represents a baseline of $\mathcal{R}_{med.}=1$ in the absence of any medium effect. }
		\vspace{-0.5cm}
		\label{fig_1}
	\end{center}
\end{figure} 

To further demonstrate this feature, let us compute the $\mathcal{R}_{med.} $ for such a simple partonic medium, essentially a QGP ``brick''~\cite{Armesto:2011ht} with constant and homogeneous temperature which extends along a given direction with a fixed width.  
In Fig.~\ref{fig_1}, the individual contributions from suppression term (blue) and enhancement term (red) as well as the overall $\mathcal{R}_{med.}$ (black) are plotted as functions of the QGP ``brick'' length, showing the nontrivial decrease-then-increase behavior of $\mathcal{R}_{med.}$ due to the competition between suppression and enhancement. This behavior already qualitatively agrees with the trends seen in experimental data. 
Of course, the partonic medium created in those collisions is much more complicated than the simple approximation here, due to nontrivial initial conditions, event-by-event fluctuations, dynamical expansions, etc. To fully verify the feasibility of this idea, one needs to perform quantitative and realistic simulations, which we report next.   

One important issue not discussed so far is the kinematic factors such as transverse momentum. The geometric path approach here is suitable for  describing a rapidly traversing probe and best applicable for extracting the integrated average medium effects on the relatively high $p_T$ region. A more differential treatment that implements initial $c\bar{c}$ $p_T$ spectrum as well as $p_T$-dependence for both suppression and enhancement effects, will be a future goal.


{\it Results.}
To quantitatively evaluate the medium effect on the $\chi_{c1} (3872)$ production relative to the $\psi(2S)$, we utilize event-by-event simulations based on the iEBE-VISHNU hydrodynamic model~\cite{Shen:2014vra} which has been well tested by experimental data from small to large colliding systems~\cite{Plumberg:2020jux, Taghavi:2019mqz, Zhao:2020pty, ATLAS:2022kqu,Shen:2015qba, Lacey:2022plw, Oliinychenko:2021enj,Wang:2020qwe, GarciaMontero:2021wyh, Dai:2018mhw, Mitrankov:2022xcr, Abbasi:2017ajp, Zhao:2018lyf, Zhu:2016qiv, Lacey:2022baf, Xu:2021uar, Heinz:2015arc, Ru:2017nkc, Chang:2015hqa, Li:2017via, Mehrabpour:2018kjs, Li:2022tcr, Zhu:2015dfa, Mordasini:2019hut, Mitrankov:2021xez, Zhu:2016puf, Aronson:2017ymv, Song:2017wtw, Fu:2015wba, ALICE:2018vhm, Devetak:2021cyq,Shen:2016egw,Bernhard:2016tnd,ALICE:2019hno,ALICE:2018yph,ALICE:2018cpu,Pang:2016vdc,Song:2017wtw,Zhao:2017yhj,ALICE:2016cti,Qian:2016fpi,Zhao:2017rgg,ALICE:2018vhm,Zhao:2018lyf,Xu:2016hmp,Du:2019civ,Shen:2015qta}.  
The iEBE-VISHNU provides event-wise time-dependent evolution information of the bulk medium as well as initial conditions for pp, pA, and AA collisions. We generate $c\bar{c}$ pairs at different spots on the event plane according to the initial binary collision density profiles. The pairs then move along straight paths whose directions are randomly chosen, and along each path the bulk medium is evolving in time. 
At each spacetime point the local entropy density can be read from the iEBE-VISHNU hydrodynamic evolution, which is directly related to active degrees of freedom in the thermal medium and thus can be used to represent the local parton density up to a proportionality constant.
In this work, we simulated a total of 500,000 events for pp collisions at $\sqrt{S_{NN}}=8$ TeV,  200,000 events for pPb collisions at $\sqrt{S_{NN}}=8.16$ TeV and 100,000 events for PbPb collisions at  $\sqrt{S_{NN}}=5.02$ TeV. Model parameters of the hydro code were set in the same way as previous studies based on the same package in the literature, see e.g. \cite{Zhu:2016puf, Li:2017via, Zhao:2020pty} .  The bulk properties such as multiplicity and elliptic flow from our simulations compare well with experimental measurements for all colliding systems.
For each event, we further simulate about 100 to 10000 in-medium paths depending on the medium size. Due to the substantial amount of needed computing time, we chose to simplify the calculations by first performing average over the path integrations in each event and then computing the exponential for the ratio between  $\chi_{c1} (3872)$ and $\psi(2S)$. That is: 
\begin{eqnarray}
\mathcal{R}_{med.} \approx  \langle e^{ -\alpha' \cdot  P_1+\beta^{'2} \cdot  P_2}\rangle_{\mathrm{event}},
\end{eqnarray}
where 
\begin{eqnarray}
P_1&=&\left\langle \int_{\mathrm{path}} s(x)\md x\right\rangle_{\mathrm{path}},\\
P_2&=&\left\langle \int_{\mathrm{path}} s(x) \left (\int_0^x s(y)\md y \right ) \md x\right\rangle_{\mathrm{path}}.
\end{eqnarray}
where we introduce $s(x)$ as the entropy density, and $\alpha'= (\alpha_X-\alpha_{\psi(2S)}) \left( \frac{n}{s}\right)$, $\beta'=\beta_X \left(\frac{n}{s}\right)$ whose definitions absorb the proportionality constant between parton density and entropy density.

To determine the two key parameters $\alpha'$ and $\beta'$, we take the LHCb pp (at $\sqrt{S_{NN}}=8$ TeV)  and preliminary pPb (at $\sqrt{S_{NN}}=8.16$TeV) as well as the CMS PbPb (at $\sqrt{S_{NN}}=5.02$ TeV) data for a global fitting analysis,  
with results shown in Fig.~\ref{fig_2}. The best fit, with $\chi^2/d.o.f=1.78$, gives the following numbers with $1\sigma$ level uncertainty: $\alpha' = (5.7\pm 2.2)\times 10^{-3}fm^2$, $\beta'=(5.7\pm0.7)\times 10^{-3}fm^2$ and $\sigma^X_{pp}/\sigma^{\psi(2S)}_{pp} = 0.162\pm 0.037$. Note the analysis does not assume the sign of $\alpha'$ which is found to be positive from the best fit.
As one can see, our model with just two parameters characterizing a competition between suppression and enhancement can well describe the quantitative trends of all global data from small to large systems. The newly proposed medium-assisted enhancement is particularly important for understanding the rapid increase of $\chi_{c1} (3872)$ yield relative to $\psi(2S)$ in the pPb and PbPb collisions. The phenomenological extraction of the two parameters from experimental data can serve as important constraints on the nature of $\chi_{c1} (3872)$. Microscopic calculations of these parameters based on different assumptions (e.g. hadronic molecule versus tetraquark state) could then be compared with the empirical values obtained here to help decipher its internal structures, which  will be an interesting future project.

\begin{figure}[!hbt]
	\begin{center}
		\includegraphics[width=3.3in]{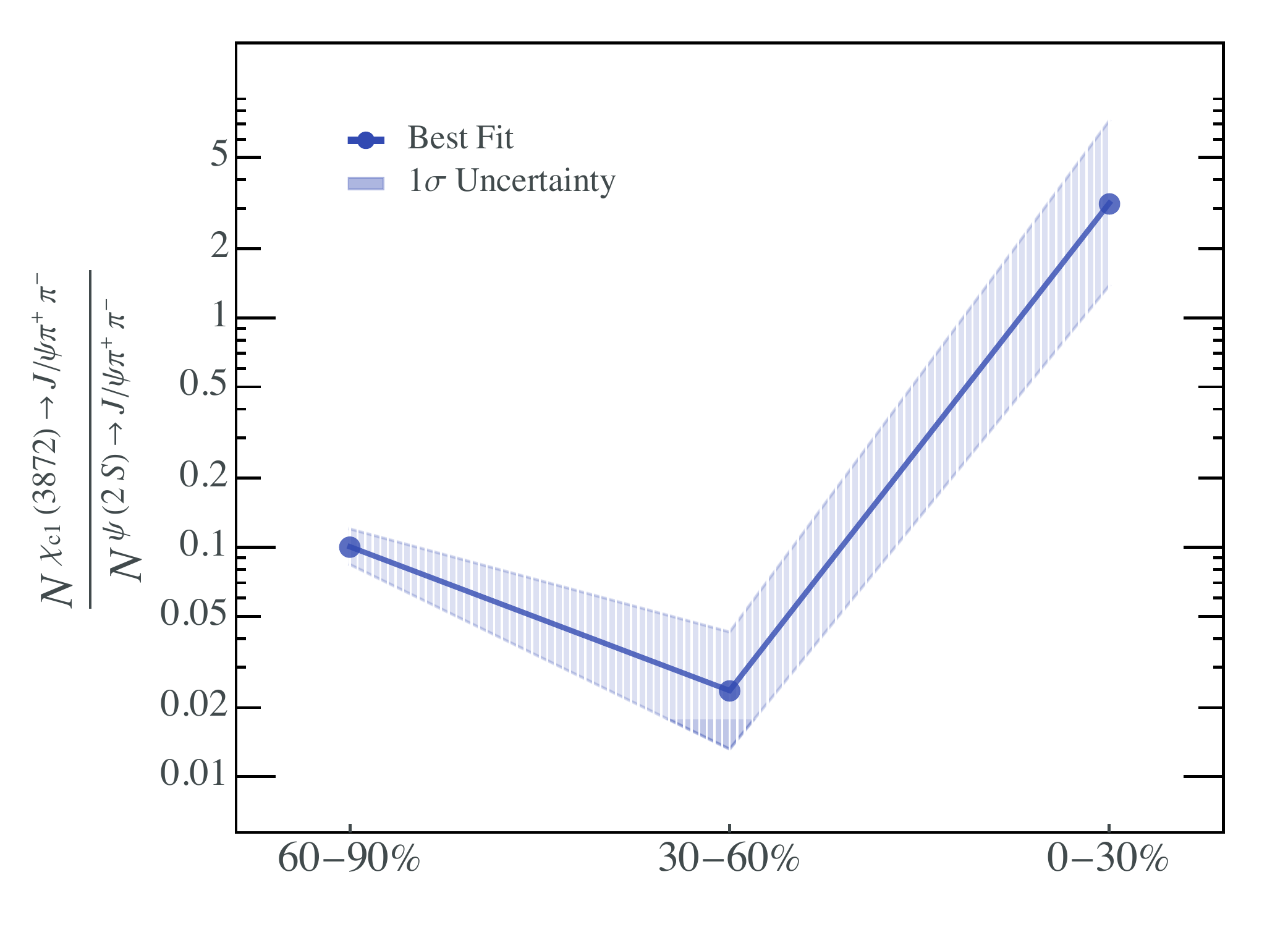}
		\caption{The predicted centrality dependence of the $\chi_{c1} (3872)$ yield relative to $\psi(2S)$ in PbPb collisions at $\sqrt{S_{NN}}=5.02$TeV collisions. The blue uncertainty band is from the same source as in Fig.~\ref{fig_2}. }
		\vspace{-0.5cm}
		\label{fig_3}
	\end{center}
\end{figure}

A natural next step would be testing our model predictions for which experimental data are not yet available and can serve as a future validation. To do that, we further investigate the centrality dependence of the medium effect in the PbPb collisions.  
In Fig~\ref{fig_3},  we show the yield ratio of $\chi_{c1} (3872)$ to $\psi(2S)$ in three centrality bins: 
$60-90\%$, $30-60\%$, and $0-30\%$. 
Interestingly, the results again show a non-monotonic behavior. In peripheral collisions the ratio is about 0.1, while in middle-centrality class it will drop to around 0.02 which is comparable to that in the pPb collisions. Finally in the central collisions, it rockets up to be as high as 3. 
This finding also suggests that the most central collisions actually contribute most of the $\chi_{c1} (3872)$ particles observed in the minimal bias measurements from CMS. We emphasize that the model parameters were already fixed in the aforementioned fitting analysis, so the highly nontrivial centrality trend predicted by the model here will be an important verification by future measurements.

\begin{figure}[!hbt]
	\begin{center}
		\includegraphics[width=3.3in]{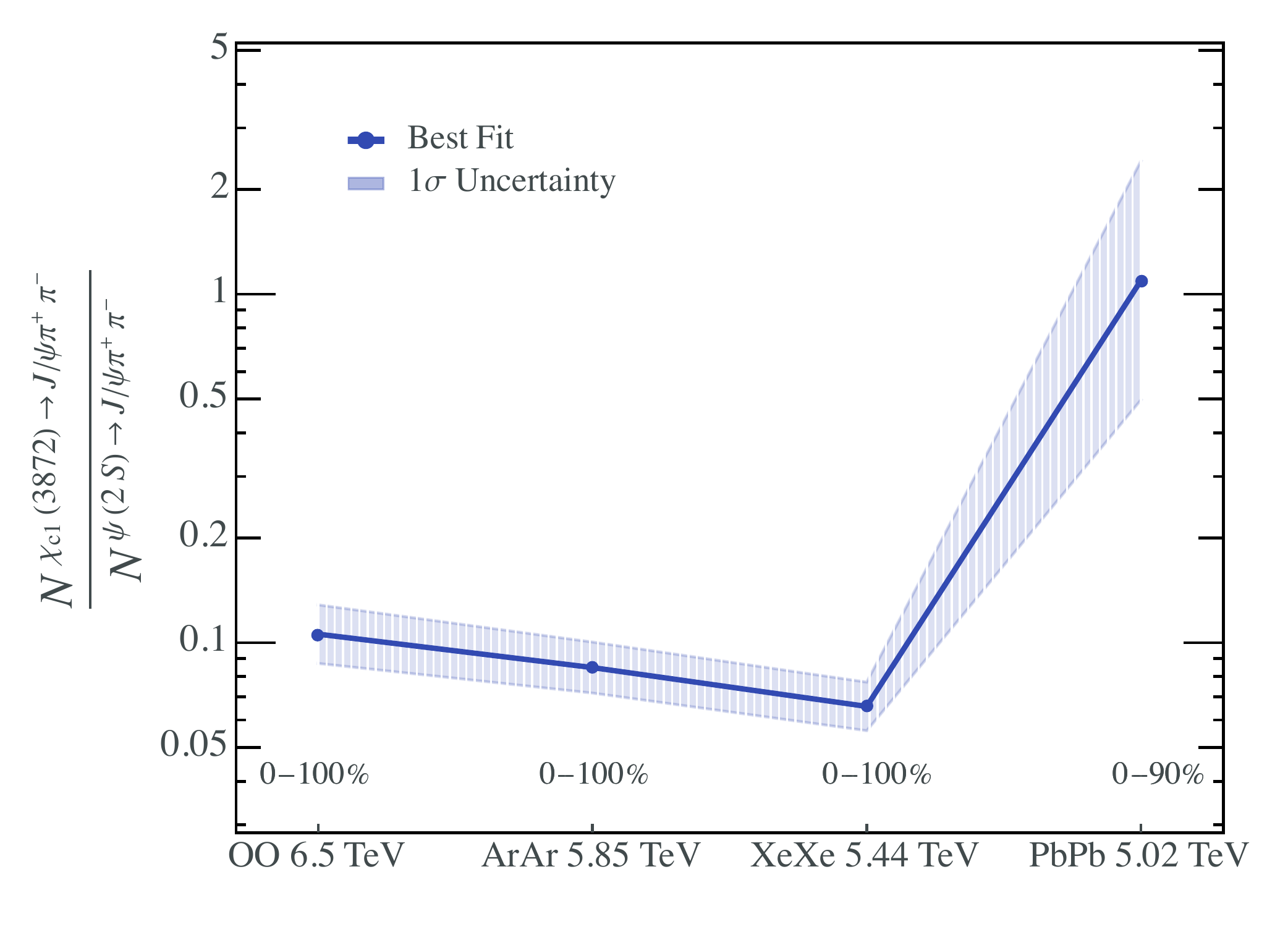}
		\caption{
		The predicted trend of the $\chi_{c1} (3872)$ yield relative to $\psi(2S)$ 
		in AA collisions from small to large systems, including: OO collisions at $\sqrt{S_{NN}}= 6.5$TeV, ArAr collisions at $\sqrt{S_{NN}}= 5.85$TeV,    
XeXe collisions at $\sqrt{S_{NN}}= 5.44$TeV, as well as  PbPb collisions at  $\sqrt{S_{NN}}= 5.02$TeV. 
		The blue uncertainty band is from the same source as in Fig.~\ref{fig_2}. }
		\vspace{-0.5cm}
		\label{fig_4}
	\end{center}
\end{figure}

Finally, the system size scan for AA collisions could offer yet another independent validation of our model predictions. For that purpose, we have computed the $\chi_{c1} (3872)$ to $\psi(2S)$ yield ratio for the following systems: 
OO collisions at $\sqrt{S_{NN}}= 6.5$ TeV, 
ArAr collisions at $\sqrt{S_{NN}}= 5.85$ TeV,   
and 
XeXe collisions at $\sqrt{S_{NN}}= 5.44$ TeV. 
These results are shown in Fig.~\ref{fig_4} in comparison with PbPb collisions at  $\sqrt{S_{NN}}= 5.02$ TeV.  Again, one observes a nontrivial trend that first decreases and then increases when changing from smaller to larger colliding systems. Such prediction of the system size dependence in AA collisions, which is the consequence of competing suppression and enhancement effects, can be readily tested with future measurements.


{\it Conclusion.}
To conclude, we present a phenomenological model for the partonic medium attenuation effects on the production of  $\chi_{c1} (3872)$ and $\psi(2S)$ particles in high energy hadron and nuclear collisions. 
In particular, a novel mechanism of medium-assisted enhancement effect is proposed for the $\chi_{c1} (3872)$ production, which leads to a competition with the more conventional absorption-induced suppression effect and becomes more dominant for higher parton densities and larger medium size. As a consequence of this important feature, the yield ratio of  $\chi_{c1} (3872)$ relative to the $\psi(2S)$ develops a nontrivial pattern, first decreasing then increasing, when the partonic medium evolves from small to large colliding systems. Utilizing realistic simulations, we show that this model offers the first quantitative description of all available experimental measurements, including  
 the LHCb pp (at $\sqrt{S_{NN}}=8$ TeV)  and preliminary pPb (at $\sqrt{S_{NN}}=8.16$TeV) as well as the CMS PbPb (at $\sqrt{S_{NN}}=5.02$ TeV) data. We further make predictions for the centrality dependence of the $\chi_{c1} (3872)$-to-$\psi(2S)$ yield ratio in PbPb collisions as well as for its system size dependence from OO and ArAr to XeXe and PbPb collisions. In both cases, a non-monotonic pattern emerges as the imprint of the competition between enhancement and suppression. Given the expected abundance of experimental data from planned runs as well as anticipated upgrades at the LHC, it would be exciting to test these predictions with future high precision measurements. 


\begin{acknowledgments}
{\it Acknowledgments.}
    This research was supported in part by the National Natural Science Foundation of China (NSFC) under Grants No.~12035007, No.~12022512 and No.~11905066, by Guangdong Major Project of Basic and Applied Basic Research No.~2020B0301030008, by the National Science Foundation in US under Grant No.~PHY-2209183 (J.L.), and by the DOE through ExoHad Topical Collaboration. 
\end{acknowledgments}

\providecommand{\noopsort}[1]{}\providecommand{\singleletter}[1]{#1}%

\end{document}